\documentstyle[epsfig,hip-artc]{article}  
\volnumber{9}  \edyear{1999}  \frompage{000} \topage{000}                %| 
\recrevdate{1 January 1999}                                              %| 
\title{HFB Calculations Near the Drip Lines} 
\authors{
{\twerm E. Ter\'an$^1$, V.E. Oberacker$^1$ and A.S. Umar $^1$ %
}\\[2.812mm]
{\normalsize
\hspace*{-8pt}$^1$ Department of Physics and Astronomy, Vanderbilt University, \\ 
Nashville, TN 37235, USA\\[0.2ex] 
}}
 
\abstract{We present the first set of results of solving the Hartree-Fock-Bogoliubov 
equations, which describe the self-consistent mean field theory with pairing 
interaction. Calculations for even-even nuclei are carried out on a 
two-dimensional axially symmetric lattice, in coordinate space. An important 
aspect of our method is the proper representation of the quasi-particle 
continuum wavefunctions, which are considered for energies up to 60 MeV. 
This stage is essential for a proper description of nuclei near the drip lines, 
due to the strong coupling between weakly bound states and the 
particle continuum for such nuclei. 
High accuracy is achieved by representing the operators and wavefunctions using 
the technique of basis-splines.
Calculations for $Sn$ isotopes are 
presented to demonstrate the reliability of the method.}

\begin{document}
 
\maketitle

\section{Introduction}

The study of structures and reactions of nuclei far from stability has been
one of the most active fields of nuclear physics in the past decade 
\cite{ISOL97,RIA1}.
The microscopic description of such nuclei will lead to a better understanding
of the interplay among the strong, Coulomb, and the weak interactions as well
as the enhanced correlations present in these many-body systems.

The experimental developments as well as recent advances in
computational physics have sparked renewed interest in nuclear
structure theory. In contrast to the well-understood behavior near the
valley of stability, there are many open questions as we move towards
the proton and neutron driplines and towards the limits in mass number. 
The neutron dripline represents mostly
``terra incognita''. In these exotic regions of the nuclear chart, some of 
the topics of interest are the effective N-N interaction at large isospin, 
large pairing correlations and their density dependence, neutron halos/skins, 
and proton radioactivity. Specifically, we are interested in calculating
ground state observables such as the total binding energy, charge radii,
proton and neutron densities, separation energies for neutrons and protons,
pairing gaps, and potential energy surfaces. It is generally acknowledged
that an accurate treatment of the pairing interaction is essential for
describing exotic nuclei \cite{DF84,DN96}.

In Section 2 the general outline of the HFB formalism is shown in second
quantization and its representation in coordinate space. Section 3 shows 
the reduction of the HFB equations for cylindrical coordinates. The last two 
sections include results and discussion of our HFB calculations.
 
\section{Hartree-Fock-Bogoliubov Formalism}  
\subsection{Basic outline of HFB equations}
The many-body Hamiltonian in occupation number representation has the form
\begin{equation}
\hat{H}= \sum_{i,j} < i|\ t\ |j> \ \hat{c}_i^\dagger \ \hat{c}_j \ +
\frac{1}{4} \sum_{i,j,m,n} <ij|\ \tilde{v}^{(2)} \ |mn>
\ \hat{c}_i^\dagger \ \hat{c}_j^\dagger \ \hat{c}_n \ \hat{c}_m	\ + \ ...
\end{equation}
\noindent Similar to the BCS theory, one performs a canonical transformation
to quasiparticle operators $\hat{\beta},\hat{\beta}^\dagger$
\begin{equation}
\left( 
\begin{array}{c}
\hat{\beta} \\ 
\hat{\beta}^\dagger 
\end{array}
\right) =
\left( 
\begin{array}{cc}
 U^\dagger & V^\dagger \\ 
  V^T & U^T 
\end{array}
\right) 
\left( 
\begin{array}{c}
\hat{c} \\ 
\hat{c}^\dagger
\end{array}
\right) \ .
\end{equation}
\noindent The HFB ground state is defined as the quasiparticle vacuum
\begin{equation}
\hat{\beta}_k \ | \Phi_0 > \ = \ 0 \ .
\end{equation}
\noindent The HFB ground state energy together with the constraint on the 
particle number is given by
\begin{equation}
E({\mathcal{R}}) = < \Phi_0 | \hat{H} - \lambda \hat{N} | \Phi_0 > \ ,
\end{equation}

\noindent where $\mathcal{R}$ is the generalized density matrix, function of the normal density $\rho_{i,j}$ 
and the pairing tensor $\kappa_{i,j}$. We derive the equations of motion from 
the variational principle
\begin{equation}
\delta \ [ E( {\mathcal{R}} ) - {\rm{tr}}\  \Lambda ( {\mathcal{R}}^2 - {\mathcal{R}} ) ]  =  0 \ ,
\end{equation}
\noindent which results in the standard HFB equations
\begin{equation}
[ {\mathcal{H}}, {\mathcal{R}} ] \ = \ 0 \ ,
\end{equation}
\noindent with the generalized single-particle Hamiltonian
\begin{equation}
{\mathcal{H}} =
\left( 
\begin{array}{cc}
 (h - \lambda) & \tilde h \\ 
 \tilde h & -(h - \lambda) \ ,
\end{array}
\right) \ ,
\end{equation}
where $h$ and $\tilde h$ denote the particle and pairing Hamiltonians, 
respectively and the Lagrange multiplier $\lambda$ is the Fermi energy.
%------------------------------------------------------------------------------
 
\subsection{HFB equations in coordinate space}

For certain types of effective interactions (e.g. Skyrme mean field and pairing
delta-interactions) the ``particle'' Hamiltonian $h$ and the
``pairing'' Hamiltonian $\tilde h$ are diagonal in isospin space and
local in position space,
\begin{equation}
h({\bf r} \sigma q, {\bf r}' \sigma' q') \ = \ \delta_{q,q'}
      \ \delta({\bf r} - {\bf r}') h^q_{\sigma, \sigma'}({\bf r}) 
\end{equation}
and
\begin{equation}
\tilde{h}({\bf r} \sigma q, {\bf r}' \sigma' q') \ = \ \delta_{q,q'}
      \ \delta({\bf r} - {\bf r}') \tilde{h}^q_{\sigma, \sigma'}({\bf r}) \ .
\end{equation}
Inserting these into the above HFB equations results in
a 4x4 structure in spin space:
\begin{equation}
\left( \matrix{ ( h^q -\lambda ) & \tilde h^q \cr 
                  \tilde h^q & - ( h^q -\lambda ) \cr} \right)
\left( \matrix{ \phi^q_{1,\alpha} \cr
                \phi^q_{2,\alpha} \cr} \right) = E_\alpha
\left( \matrix{ \phi^q_{1,\alpha} \cr
                \phi^q_{2,\alpha} \cr} \right) \ ,
\label{hfbeq2}
\end{equation}
with
\begin{equation}
h^q = \left( \matrix{ h^q_{\uparrow \uparrow}({\bf r}) & h^q_{\uparrow \downarrow}({\bf r}) \cr 
                  h^q_{\downarrow \uparrow}({\bf r}) & h^q_{\downarrow \downarrow}({\bf r}) \cr} \right) \ , \ \ \ 
\tilde h^q = \left( \matrix{ \tilde h^q_{\uparrow \uparrow}({\bf r}) & \tilde h^q_{\uparrow \downarrow}({\bf r}) \cr 
                  \tilde h^q_{\downarrow \uparrow}({\bf r}) & \tilde h^q_{\downarrow \downarrow}({\bf r}) \cr} \right)
\label{hspin}
\end{equation}
and
\begin{equation}
\phi^q_{1,\alpha} = \left( \matrix{ \phi^q_{1,\alpha}({\bf r},\uparrow) \cr
                \phi^q_{1,\alpha}({\bf r},\downarrow) \cr} \right) \ , \ \ \ 
\phi^q_{2,\alpha} = \left( \matrix{ \phi^q_{2,\alpha}({\bf r},\uparrow) \cr
                \phi^q_{2,\alpha}({\bf r},\downarrow) \cr} \right) \ .
\label{phispin}
\end{equation}

Because of the
structural similarity between the Dirac equation and the HFB equation in
coordinate space, we encounter here similar computational challenges: for
example, the spectrum of quasiparticle energies $E$ is unbounded from above {\em
and} below. The spectrum is discrete for $|E|<-\lambda$
and continuous for $|E|>-\lambda$. For even-even nuclei it is customary to 
solve the HFB equations with a positive
quasiparticle energy spectrum $+E_\alpha$ \cite{RS80} and consider all negative
energy states as occupied in the HFB ground state.

%------------------------------------------------------------------------------
\section{2-D Reduction for Axially Symmetric Systems}

For simplicity, we assume that the HFB quasi-particle Hamiltonian
is invariant under rotations ${\hat R}_z$ around
the z-axis, i.e. $[{\mathcal{H}},{\hat R}_z]=0$. Due to the axial symmetry
of the problem, it is advantageous to introduce cylindrical coordinates
$(\phi,r,z)$.

It is possible to construct simultaneous
eigenfunctions of the generalized Hamiltonian ${\mathcal{H}}$ and
the z-component of the angular momentum, ${\hat j}_z$
\begin{eqnarray}
{\mathcal{H}} \ \psi_{n,\Omega,q} (\phi,r,z) & = & E_{n,\Omega,q} \ 
    \psi_{n,\Omega,q} (\phi,r,z)  \nonumber   \\
{\hat j}_z \ \psi_{n,\Omega,q} (\phi,r,z) & = & \hbar \Omega \ 
    \psi_{n,\Omega,q} (\phi,r,z) \ ,
\end{eqnarray}
with the quantum numbers 
$\Omega = \pm \frac{1}{2}, \pm \frac{3}{2},\pm \frac{5}{2}, ...$.
The simultaneous quasiparticle eigenfunctions take the form
\begin{equation}
\psi_{n,\Omega,q} (\phi,r,z) = 
\left( 
\begin{array}{c}
\phi^{(1)}_{n,\Omega,q} (\phi,r,z) \\ 
\phi^{(2)}_{n,\Omega,q} (\phi,r,z) \\ 
\end{array} 
\right)
= \frac{1}{\sqrt{2 \pi}}
\left( 
\begin{array}{c}
e^{i(\Omega - \frac 12)\phi} \ \phi^{(1)}_{n,\Omega,q} (r,z,\uparrow) \\ 
e^{i(\Omega + \frac 12)\phi} \ \phi^{(1)}_{n,\Omega,q} (r,z,\downarrow) \\
e^{i(\Omega - \frac 12)\phi} \ \phi^{(2)}_{n,\Omega,q} (r,z,\uparrow) \\ 
e^{i(\Omega + \frac 12)\phi} \ \phi^{(2)}_{n,\Omega,q} (r,z,\downarrow)
\end{array} 
\right) \ .
\label{eq:wvfnctn}
\end{equation}
We introduce the following useful notation
\begin{equation}
U^{(1,2)}_{ n \Omega q} (r,z) =  \phi^{(1,2)}_{n,\Omega,q} (r,z,\uparrow)\ , \  
L^{(1,2)}_{ n \Omega q} (r,z) =  \phi^{(1,2)}_{n,\Omega,q} (r,z,\downarrow) \ .
\end{equation}

From the vanishing commutator, $ [ {\mathcal H}, j_z ]$, we
can determine the $\phi$-dependence of the HFB quasi-particle Hamiltonian
and arrive at the following structure for the Hamiltonian
\begin{equation}
h (\phi,r,z) =
\left( 
\begin{array}{cc}
h^{\prime}_{\uparrow \uparrow} \ (r,z) & e^{-i \phi} \ h^{\prime}_{\uparrow \downarrow} \ (r,z) \\ 
e^{+i \phi} \ h^{\prime}_{\downarrow \uparrow} \ (r,z) & h^{\prime}_{\downarrow \downarrow} \ (r,z)
\end{array}
\right) \ .
\label{eq:hamil}
\end{equation}
and the pairing Hamiltonian
\begin{equation}
\tilde h (\phi,r,z) =
\left( 
\begin{array}{cc}
\tilde h^{\prime}_{\uparrow \uparrow} \ (r,z) & e^{-i \phi} \ \tilde h^{\prime}_{\uparrow \downarrow} \ (r,z) \\ 
e^{+i \phi} \ \tilde h^{\prime}_{\downarrow \uparrow} \ (r,z) & \tilde h^{\prime}_{\downarrow \downarrow} \ (r,z)
\end{array}
\right) \ ,
\label{eq:hamil-tilda}
\end{equation}
Inserting equations (\ref{eq:hamil}) and  (\ref{eq:hamil-tilda}) into the eigenvalue Eq. (\ref{hfbeq2})
, we arrive at the ``reduced 2-D problem'' \cite{Obe99}
in cylindrical coordinates:
\begin{eqnarray}
\left( \matrix{ (h'_{\uparrow \uparrow} - \lambda) & h'_{\uparrow \downarrow} &
                   \tilde{h'}_{\uparrow \uparrow} & \tilde{h'}_{\uparrow \downarrow} \cr
               h'_{\downarrow \uparrow} & (h'_{\downarrow \downarrow} - \lambda) &
	           \tilde{h'}_{\downarrow \uparrow} & \tilde{h'}_{\downarrow \downarrow} \cr
	       \tilde{h'}_{\uparrow \uparrow} & \tilde{h'}_{\uparrow \downarrow} &
	            -(h'_{\uparrow \uparrow} - \lambda) &  -h'_{\uparrow \downarrow}\cr
               \tilde{h'}_{\downarrow \uparrow} & \tilde{h'}_{\downarrow \downarrow} &
	             -h'_{\downarrow \uparrow} & -(h'_{\downarrow \downarrow} - \lambda) \cr} \right)
\left( \matrix{ U^{(1)}_{n,\Omega,q} \cr
                L^{(1)}_{n,\Omega,q} \cr
                U^{(2)}_{n,\Omega,q} \cr
                L^{(2)}_{n,\Omega,q} \cr} \right) = E_{n,\Omega,q} \ 
\left( \matrix{ U^{(1)}_{n,\Omega,q} \cr
                L^{(1)}_{n,\Omega,q} \cr
                U^{(2)}_{n,\Omega,q} \cr
                L^{(2)}_{n,\Omega,q} \cr} \right)
\label{eq:hfb2d}\nonumber
\end{eqnarray}

Here $\tilde{h'}$, $h'$, $U$'s, and $L$'s are all functions of $(r,z)$ only.
For a given angular momentum projection quantum number $\Omega$, we solve
the eigenvalue problem to obtain energy 
eigenvalues $E_{n,\Omega,q}$ and eigenvectors $\psi_{n,\Omega,q}$
for the HFB quasi-particle states. 

%------------------------------------------------------------------------------
\subsection{HFB Hamiltonian using the Skyrme interaction}

Using the Skyrme form for the particle Hamiltonian, $h_q$, we can write
\begin{equation}
\label{hf_hamiltonian}
h_q = - {\bf\nabla  \cdot} \frac{\hbar^2}{2 m_q^*} {\bf \nabla} 
    + U_q + U_C
    - i {\bf B}_q{\bf \cdot}\left( {\bf \nabla \times \sigma} \right)\;\;.
\end{equation}
where the effective mass is defined by   
\begin{equation}
\label{effective_mass_def}
\frac{\hbar^2}{2 m_q^*} = \frac{\hbar^2}{2m_q} 
 + \frac{1}{4} (t_1 +t_2) \rho  + \frac{1}{8}(t_2 - t_1) \rho_q\;\;.
\end{equation}
Applying the cylindrical form of the Laplacian operator to the standard
form of the wavefunction in Eq.(\ref{eq:wvfnctn}), 
and invoking the axial symmetry of $f$ we find

\begin{equation}
\hat{t}_q = \left( \begin{array}{cc} t_{11} & 0 \\ 0 & t_{22} \end{array} \right)
\end{equation}
where the elements are given by
\begin{eqnarray}
t_{11} &=& 
 f \left( \frac{\partial^2}{\partial r^2} 
 + \frac{1}{r} \frac{\partial}{\partial r}
 - \left(\frac{(\Omega - 1/2)}{r}\right)^2 
 + \frac{\partial^2}{\partial z^2} \right) 
 + \frac{\partial f}{\partial r} \frac{\partial}{\partial r}
 + \frac{\partial f}{\partial z} \frac{\partial}{\partial z} \\
t_{22} &=& 
 f \left( \frac{\partial^2}{\partial r^2} 
 + \frac{1}{r} \frac{\partial}{\partial r}
 - \left(\frac{(\Omega + 1/2)}{r}\right)^2 
 + \frac{\partial^2}{\partial z^2} \right) 
 + \frac{\partial f}{\partial r} \frac{\partial}{\partial r}
 + \frac{\partial f}{\partial z} \frac{\partial}{\partial z} \ ,
\end{eqnarray}
$f$ being the effective mass.
The local potential terms could also be cast into a matrix form
\begin{equation}
\hat{v}_q = \left( \begin{array}{cc} v_{11} & 0 \\ 0 & v_{22} \end{array} \right) \ ,
\end{equation}
where 
\begin{equation}
v_{11} = v_{22} = U_q + U_C\;\;.
\end{equation}
The individual terms are constructed simply by summing appropriately weighted 
scalars as indicated by
Eqs.~(\ref{effective_potential}) and including the Coulomb 
potential and the Slater exchange (\ref{coulomb_potential}) term: 
\begin{eqnarray}
\label{effective_potential}
U_q &=& t_0\left[ \left( 1 + \frac{1}{2} x_0\right) \rho - 
                   \left( \frac{1}{2} + x_0 \right) \rho_q \right] \nonumber \\
      &+& \frac{1}{4} \left(t_1 + t_2 \right) \tau - 
          \frac{1}{8}\left(t_1 - t_2 \right) \tau_q \nonumber \\
      &+& \frac{1}{12} t_3 \left(1 + \frac{1}{2}x_3\right)
          \left( \alpha + 2 \right) \rho^{\alpha + 1} \\
      &-& \frac{1}{12} t_3 \left(\frac{1}{2} + x_3\right)
          \left[ \alpha \rho^{\alpha - 1} \sum_q \rho_q^2 
          + 2 \rho^{\alpha} \rho_q  \right] \nonumber \\
      &-& \frac{1}{2} t_4 \left[ {\bf \nabla \cdot J + \nabla \cdot J}_q
      \right] \nonumber \\
      &+& \frac{1}{8}\left(t_2 - 3 t_1\right)\nabla^2 \rho +
      \frac{1}{16} \left( t_2 + 3 t_1 \right) \nabla^2 \rho_q \nonumber \;\;,
\end{eqnarray}
\begin{equation}
\label{coulomb_potential}
U_C = e^2 \int d^3 r' \frac{\rho_p({\bf r'})}{|{\bf r} - {\bf r'}|} 
    - e^2 \left(\frac{3}{\pi} \right)^{1/3}\left[\rho_p({\bf r})\right]^{1/3}\;\;.
\end{equation}
The Hartree-Fock spin-orbit operator
\begin{equation}
\label{hf_ls} 
- i {\bf B}_q{\bf \cdot}\left( {\bf \nabla \times \sigma} \right) 
\longrightarrow   \hat{w}_q\;\;,
\end{equation}
could similarly be cast into the form
\begin{equation}
\hat{w}_q  = \left( \begin{array}{cc} w_{11} & w_{12} \\  
                           w_{21} & w_{22} \end{array} \right) \ ,
\end{equation}
with
\begin{eqnarray*}
w_{11} &=& {\cal B}_r \frac{\Omega - 1/2}{r} \\
w_{12} &=& \left[- {\cal B}_z\frac{\Omega+1/2}{r}  -
        {\cal B}_z\frac{\partial}{\partial r} + 
        {\cal B}_r \frac{\partial}{\partial z}\right]  \\
w_{21} &=& \left[-{\cal B}_z \frac{\Omega - 1/2}{r} + 
        {\cal B}_z\frac{\partial}{\partial r} 
        -{\cal B}_r \frac{\partial}{\partial z}\right] \\
w_{22} &=&  - {\cal B}_r\frac{\Omega + 1/2}{r}  \ ,             
\end{eqnarray*}
where
\begin{equation}
{\cal B}_r \equiv {\bf B_q \cdot e_r} 
= \nabla_r \frac{t_4}{2}(\rho + \rho_q) \ \  \ , \ \ \ 
{\cal B}_z \equiv {\bf B_q \cdot e_z}
= \nabla_z \frac{t_4}{2}(\rho + \rho_q)\ \ . 
\end{equation}

%------------------------------------------------------------------------------

\subsection{Densities and currents}
While the form of the single particle Hamiltonian remains the same as Skyrme HF 
Hamiltonian, the densities and currents need to be written in terms of the 
quasi-particle wavefunctions. We obtain the following
expressions for the normal density $\rho_q({\bf r})$ and pairing density
$\tilde \rho_q({\bf r})$, which are defined as the spin-averaged diagonal
elements
\begin{equation}
\rho_q({\bf r}) \ = \ \sum_{\sigma} \rho ({\bf r} \sigma q, {\bf r} \sigma q) \ = \ \sum_{\sigma}
\sum_{\alpha} \phi_{2,\alpha} ({\bf r} \sigma q) \ \phi_{2,\alpha}^* ({\bf r} \sigma q)
\label{eq:rhonormal}
\end{equation}
and
\begin{equation}
\tilde{\rho_q}({\bf r}) \ = \ \sum_{\sigma} \tilde{\rho} ({\bf r} \sigma q, {\bf r} \sigma q) \ = \ - \sum_{\sigma}
\sum_{\alpha} \phi_{2,\alpha} ({\bf r} \sigma q) \ \phi_{1,\alpha}^* ({\bf r} \sigma q) \ .
\label{eq:rhotilde}
\end{equation}
The physical interpretation of $\tilde{\rho_q}$ has been discussed in \cite{DN96}:
the quantity $[\tilde{\rho_q}({\bf r})\ \Delta V /2]^2$ gives the probability to find a
\emph{correlated} pair of nucleons with opposite spin projection in the volume
element $\Delta V$. 

Using the structure of the bi-spinor wavefunctions defined earlier
we find the following expressions for the normal and pairing densities.
\begin{equation}
  \rho_q(r,z) = \frac{1}{2 \pi} 
  \left(2 \sum_{\Omega>0}^{\Omega_{max}} \right) \sum_{E_n>0}^{E_{max}} 
  \left[|U^{(2)}_{n \Omega q}(r,z)|^2 + |L^{(2)}_{n \Omega q}(r,z)|^2 \right]  \\
\end{equation}
\begin{equation}
\tilde{\rho}_q(r,z) = - \frac{1}{2 \pi} 
  \left(2 \sum_{\Omega>0}^{\Omega_{max}} \right) \sum_{E_n>0}^{E_{max}} 
  \left[U^{(2)}_{n \Omega q}(r,z) U^{(1)*}_{n \Omega q}(r,z)
   + L^{(2)}_{n \Omega q}(r,z) L^{(1)*}_{n \Omega q}(r,z) \right] \ .
\end{equation}
Similarly, the kinetic energy density and the divergence of the current become
\begin{eqnarray}
\tau_q({\bf r}) \ = \ \nabla \cdot \nabla' \rho_q({\bf r},{\bf r}') |_{{\bf r}={\bf r}'} \ 
= \ \sum_{\sigma} \sum_{\alpha} | \nabla \ \phi_{2,\alpha} ({\bf r} \sigma q) |^2 \ , 
\label{eq:defkin}
\end{eqnarray}

\begin{eqnarray}
 \tau_q(r,z) = \frac{1}{2 \pi}
 \left(2 \sum_{\Omega>0}^{\Omega_{max}} \right) \sum_{E_n>0}^{E_{max}}
 \left[
  \frac{(\Omega - 1/2)^2}{r^2} \left| U^{(2)}_{ n \Omega q} \right|^2
  +\frac{(\Omega + 1/2)^2}{r^2} \left| L^{(2)}_{ n \Omega q} \right|^2 \right.
  \nonumber \\    
\left.  +\left| \frac{\partial U^{(2)}_{ n \Omega q}}{\partial r}  \right|^2
  +\left| \frac{\partial L^{(2)}_{ n \Omega q}}{\partial r}  \right|^2
  +\left| \frac{\partial U^{(2)}_{ n \Omega q}}{\partial z}  \right|^2
  +\left| \frac{\partial L^{(2)}_{ n \Omega q}}{\partial z}  \right|^2 \right]
\end{eqnarray}

\begin{eqnarray*}
\lefteqn{\nabla \cdot {\bf J}_q( {\bf r} )  = 
\frac{1}{2 \pi}
\left(2 \sum_{\Omega>0}^{\Omega_{max}} \right) \sum_{E_n>0}^{E_{max}}
  2  \left[
\frac{\partial U^{(2)}_{n \Omega q}}{\partial r} 
\frac{\partial L^{(2)}_{n \Omega q}}{\partial z}
- \frac{\partial L^{(2)}_{n \Omega q}}{\partial r} 
\frac{\partial U^{(2)}_{n \Omega q}}{\partial z}  \right.} \\
\;\; & &
+ \left. \frac{\Omega - 1/2}{r} U^{(2)}_{n \Omega q} 
\left( \frac{\partial U^{(2)}_{n \Omega q}}{\partial r} -
 \frac{\partial L^{(2)}_{n \Omega q}}{\partial z}  \right)
-\frac{\Omega + 1/2}{r} L^{(2)}_{n \Omega q} 
\left( \frac{\partial U^{(2)}_{n \Omega q}}{\partial z} +
 \frac{\partial L^{(2)}_{n \Omega q}}{\partial r}  \right)  
\right ] .
\end{eqnarray*}
The total number of protons and neutrons is obtained by integrating over their
densities
\begin{equation}
N_q = \int d^3r \ \rho_q({\bf r}) = 2 \pi \int_0^{\infty} r dr
      \int_{-\infty}^{\infty} dz \ \rho_q(r,z)
\end{equation}
Finally, we state the normalization condition for the four-spinor quasiparticle wavefunctions
as
\begin{equation}
\int d^3r  \ \psi^\dagger_{n \Omega q}({\bf r}) \ \psi_{n \Omega q}({\bf r}) \ = \ 1
\end{equation}
which leads to 
\begin{eqnarray*}
\int_0^{\infty} r dr \int_{-\infty}^{\infty} dz
\left[|U^{(1)}_{n \Omega q}(r,z)|^2 + |L^{(1)}_{n \Omega q}(r,z)|^2 +       
      |U^{(2)}_{n \Omega q}(r,z)|^2 + |L^{(2)}_{n \Omega q}(r,z)|^2 \right]        
\ = \ 1 \ .
\end{eqnarray*}

%------------------------------------------------------------------------------

\subsection{Pairing interaction.}
If one assumes that the effective interaction $\bar{v}^{(2)}_{pair}$ is local,
\begin{eqnarray*}
\bar{v}^{(2)}_{pair} ({\bf r} \sigma, {\bf r}' - \sigma'; {\bf r_1}' \sigma_1', {\bf r_2}' - \sigma_2') \ =
\ \delta({\bf r_1}' - {\bf r}) \ \delta_{\sigma_1',\sigma}
\ \delta({\bf r_2}' - {\bf r}') \ \delta_{\sigma_2',\sigma'}
V_p({\bf r} \sigma, {\bf r}' - \sigma')
\end{eqnarray*}
one finds the following expression for the pairing mean field
\begin{equation}
\tilde{h}({\bf r} \sigma, {\bf r}' \sigma') \ = \ V_p({\bf r} \sigma, {\bf r}' - \sigma') \ 
\tilde{\rho} ({\bf r} \sigma, {\bf r}' \sigma') \ .
\end{equation}
For the local pairing interaction $V_p$ we utilize
\begin{equation}
V_p({\bf r} \sigma, {\bf r}' - \sigma') \ = \ V_0 \ \delta({\bf r} - {\bf r}')
            \ \delta_{\sigma,\sigma'} \ F({\bf r}) \ .
\end{equation}
This parameterization describes two primary pairing forces: 
a pure delta interaction ($F=1$) that gives rise to {\it volume pairing}, and a 
density dependent delta interaction (DDDI) that gives rise to {\it surface pairing}. 
The DDDI interaction generates the following pairing mean field for the two
isospin orientations $q = \pm \frac{1}{2}$
\begin{equation}
\tilde{h}_q({\bf r} \sigma, {\bf r}' \sigma') \ = \ \frac{1}{2} \ V_0^{(q)} 
         \tilde{\rho_q}({\bf r}) F({\bf r})
         \ \delta({\bf r} - {\bf r}') \ \delta_{\sigma,\sigma'} \;\; .
\end{equation}
The pairing contribution to the nuclear binding energy is
\begin{equation}
E_{pair} = E_{pair}^{(p)} + E_{pair}^{(n)} = \int d^3 r
    \left[ \frac{V_0^{(p)}}{4} \tilde{\rho}_p^{\ 2} ({\bf r}) 
  + \frac{V_0^{(n)}}{4} \tilde{\rho}_n^{\ 2} ({\bf r}) \right] F({\bf r}) \ .
\end{equation}

%------------------------------------------------------------------------------
\subsection{Lattice representation of spinor wavefunctions and Hamiltonian}
For a given angular momentum projection quantum number $\Omega$, we solve the
eigenvalue problem on a 2-D grid $(r_\alpha,z_\beta)$ where $\alpha = 1,...,N_r$
and $\beta = 1,...,N_z$. The four components of the spinor wavefunction $\psi(r,z)$ are represented on
the 2-D lattice by a product of Basis Spline functions $B_i (x)$ evaluated at
the lattice support points. Further details are given in Ref. \cite{K96}. 

For the lattice representation of the Hamiltonian,
we use a hybrid method \cite{KO96,K96} in which
derivative operators are constructed using the Galerkin method; this 
amounts to a global error reduction. Local potentials are
represented by the B-Spline collocation method (local error reduction).
The lattice representation transforms the differential operator
equation into a matrix form
\begin{equation}
\sum_{\nu=1}^N {\cal{H}}_{\mu}^{\ \nu} \psi^{\Omega}_{\nu} = 
            E^{\Omega}_{\mu} \psi^{\Omega}_{\mu} \ \ \ (\mu=1,...,N) \ ,
\end{equation}
with $N=4 \times N_r \times N_z$.
The method of direct diagonalization with LAPACK is implemented to solve this
eigenvalue problem.
Our HFB code is written in Fortran 90 and makes extensive use of new
data concepts, dynamic memory allocation and pointer variables.
The code uses as a starting point the result of a {\it HF+BCS} calculation,
which makes HFB converge substantially faster.

Since the problem is self-consistent we use an iterative method for the
solution. The Fermi level, $\lambda$, is calculated in every iteration by means of a
simple root search, and used for the next iteration. This process is done 
until a suitable convergence is achieved.
The quasiparticle energies and corresponding wavefunctions are calculated 
up to 60 MeV. In practice a cutoff at this energy is imposed, but this limit 
can be set higher if necessary. Further details will be given elsewhere
\cite{Teran01}.

%------------------------------------------------------------------------------
\section{Results}
In table \ref{table1} we diplay the results of calculations for two tin isotopes
$^{120} Sn$ and $^{150} Sn$.
In the calculations of $^{150} Sn$ we used a box size ($R_{box}$) of 20 $fm$. The numerical 
mesh includes 17 and 34 points in $r$ and $z$ direction respectively. These points are 
geometrically distributed, giving more data points in the 
central region, where the particle and pairing densities are denser. The maximum 
$\Omega$ number was $\frac{17}{2}$ for this case.
For the 1-D calculations, $R_{box} = 30 fm$, linear spacing of 0.25 
$fm$ and $j_{max}$ of $\frac{21}{2}$ was used.

\begin{table}[htb] 
\vspace*{-12pt}
\caption[]{
Comparison of calculations for tin isotopes. The 1-D 
calculations were made by Dobaczewski \cite{DN96} using the $SkP$ 
interaction for $^{120} Sn$, and $SkM^*$ for $^{150} Sn$.
Calculations made by our HFB 2-D code used $SkM^*$.}
\vspace*{-14pt}
\begin{center}
\begin{tabular}{| l | r | r | r | r | r | }
\hline\\[-10pt]
  & \multicolumn{3}{c |}{ $^{120} Sn$ }   &   \multicolumn{2}{c |}{   $^{150} Sn$   }    \\  \cline{2-6}
  \emph{Observables} &  \emph{Exp.}   & \emph{1-D}  &  \emph{2-D}  &  \emph{1-D}  &  \emph{2-D} \\ 
\hline\\[-10pt]
Binding Energy   (Mev)     & -1020 &       & -1021      &  -1155   &  -1157  \\
Fermi Energy (n) (Mev)     &       & -7.94 &  -8.39     &  -1.86   &  -1.99  \\ 
Fermi Energy (p) (Mev)     &       &       &  -7.58     &  -16.13  & -15.95 \\ 
Pairing Gap  (n) (Mev)     & 1.378 & 1.256 &  1.85      &   1.55   &   2.10  \\ 
Pairing Gap  (p) (Mev)     &       & 0.00  &  0.00      &   0.00   &   0.00 \\ 
Beta 2                     &-0.004 &       & -0.008     &          & -0.001 \\
\hline 
\end{tabular}
\end{center}
\label{table1}
\end{table}
According to the results shown in table \ref{table1} for $Sn$ isotopes, the agreement
is evident with respect to the 1-D calculations by Dobaczewski, although $^{120} Sn$
calculations were done with different Skyrme forces. We encounter some 
differences in the pairing gap. This is likely to be explained by the dependence of 
the pairing gap on the box size, which was 30 $fm$ for 1-D calculations but 
20 $fm$ for 2-D calculations.

\begin{figure}[htb]
\vspace*{-1.0cm}
\begin{center}
\includegraphics[scale=0.27]{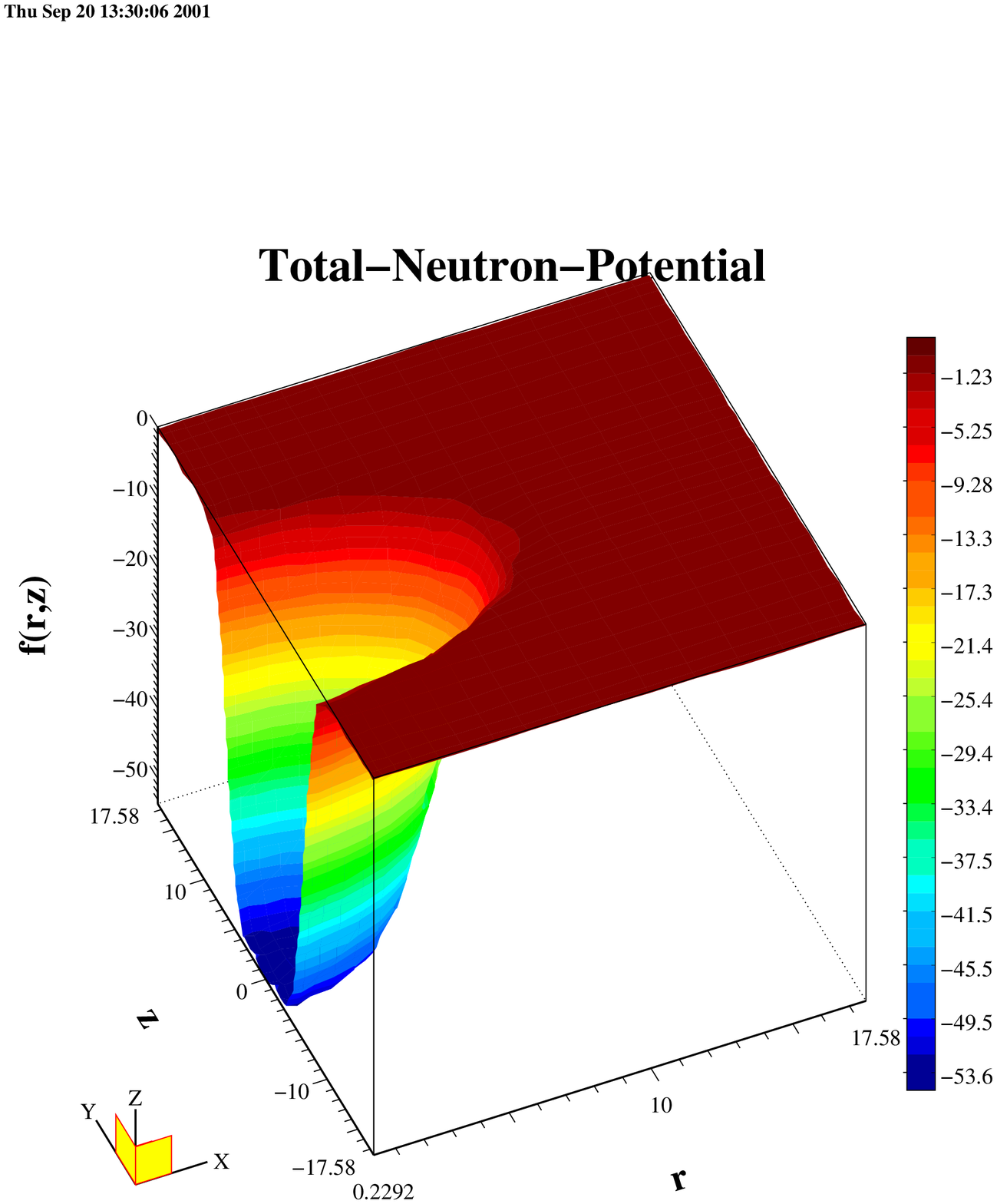}
\includegraphics[scale=0.27]{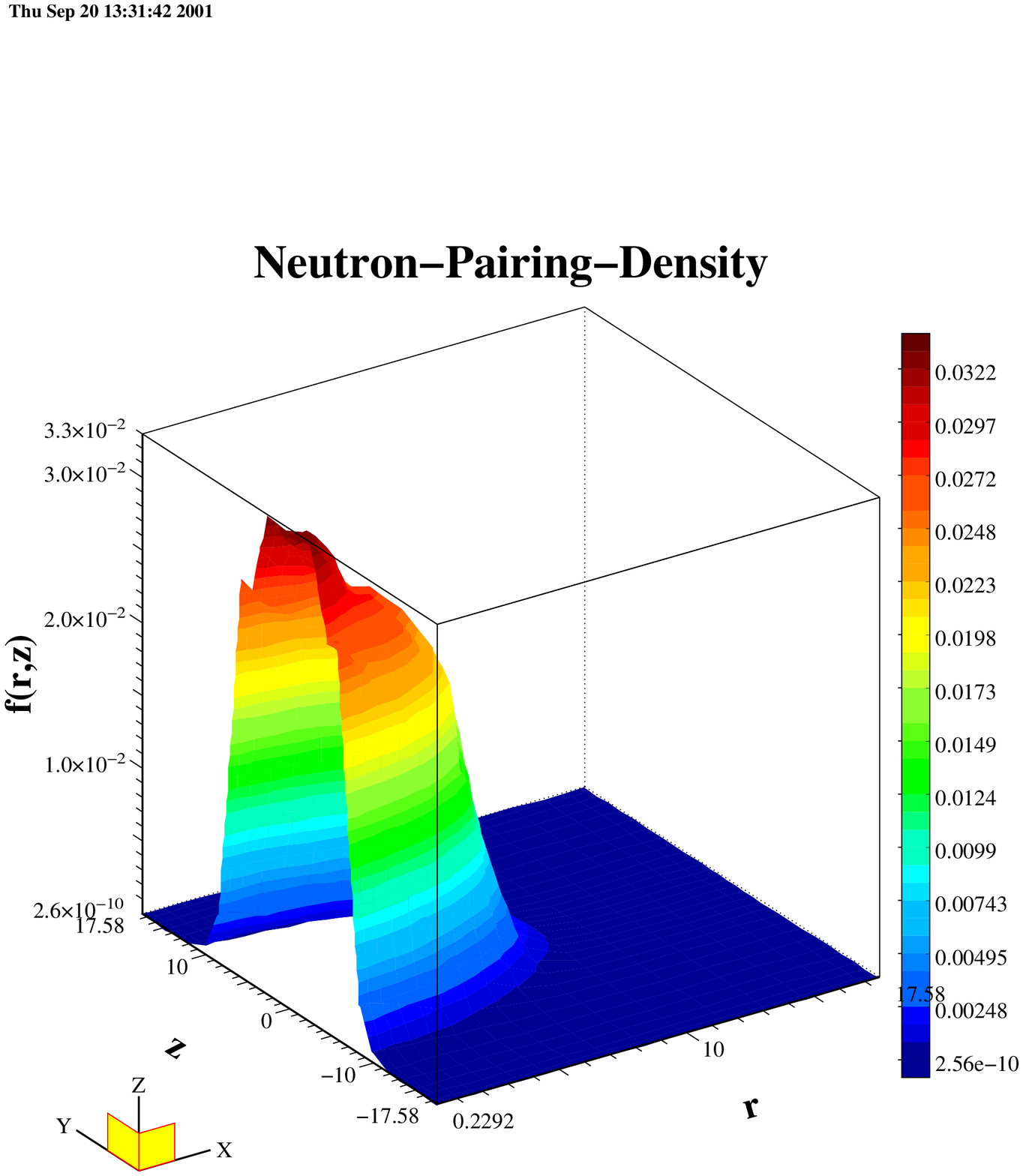}
\end{center}
\vspace*{-2.0cm}
\caption{Neutron potential and pairing density for $^{150}Sn$.}
\label{150sn_dens}
\end{figure}

In figure \ref{150sn_dens} we show the total neutron potential and the neutron 
pairing density. As we can see the pairing effect in the case of $^{150} Sn$ is 
appreciable.

\section{Conclusions}
We have seen that our calculations are close to the experimental values, and that
our two-dimensional, axially symmetric 
code agrees with the calculations of the one-dimensional HFB code \cite{DN96} for 
spherical nuclei.

The axial symmetry imposed in our HFB code is expected to be well suited in 
describing deformed nuclei far from stability. Our HFB approach 
for such nuclei works especially well in treating strong coupling 
to the continuum, which was shown to be crucial for obtaining convergence.

\section*{Acknowledgement}
This work was partially supported under U.S. Department of Energy grant
DE-FG02-96ER40963 to Vanderbilt University.

\vfill\eject

\end{document}